# INSTANTANEOUS, NON-SQUEEZED, NOISE-BASED LOGIC

FERDINAND PEPER [a] and LASZLO B. KISH [b]

[a] National Institute of Information and Communications Technology, Kobe, 651-2492 Japan

[b] Texas A&M University, Department of Electrical and Computer Engineering, College Station, TX 77843-3128, USA

**Abstract:** Noise-based logic, by utilizing its multidimensional logic hyperspace, has significant potential for low-power parallel operations in beyond-Moore-chips. However universal gates for Boolean logic thus far had to rely on either time averaging to distinguish signals from each other or, alternatively, on squeezed logic signals, where the logic-high was represented by a random process and the logic-low was a zero signal. A major setback is that squeezed logic variables are unable to work in the hyperspace, because the logic-low zero value sets the hyperspace product vector to zero. This paper proposes Boolean universal logic gates that alleviate such shortcomings. They are able to work with non-squeezed logic values where both the high and low values are encoded into nonzero, bipolar, independent random telegraph waves. Non-squeezed universal Boolean logic gates for spike-based brain logic are also shown. The advantages vs. disadvantages of the two logic types are compared.

## 1. Introduction

Noise-based informatics has received increasing attention in recent years. In noise-based logic [1-6], the information is carried by the noise, while in noise-driven information systems [7-10], noise is used to drive the system. In noise-based communication, the randomness of noise is utilized for secure key exchange [11-17].

*Instantaneous noise based logic* (*INBL*) [1-3] is a noise-based logic scheme [1-6] in which the logic 0 and 1 signals are represented by independent noise sources, whereby operations on the signals do not involve time-averaging. Unlike the noise-based schemes in [4,5], where the noise sources are continuously valued and time-averaging is intensively employed in the input correlators of gates, *INBL* relies on so-called Random-Telegraph Waves (*RTW*) [1,6] or random neural spike sequences [2,3]. *RTW*s are random square waves, valued -1 or +1, both with probabilities 0.5 at discrete points in time [1,6]. *INBL* schemes like in [1] thus allow fast logic operations without time averaging. Similarly to the continuum noise based logic schemes [4,5], signals represented by *RTW*s can be combined into hyperspace vectors (see [6] for an application) by multiplication. In the brain logic scheme [2,3], which is also *INBL* [1], the hyperspace is combined by the addition of spike sequences [2,3].

The Boolean *INBL* schemes proposed in [1] have an important problem in the formation of hyperspace: only the logic *High* (*H*) value is represented by *RTW*s, while the logic *Low* (*L*) value is squeezed to zero, out of necessity to have a useful AND gate logical operation. This squeezing of waves causes zero vectors to appear in the product hyperspace of *RTW*s, and that fact prohibits the direct use of combination of such *L* logic values in a hyperspace product containing several bits because it would reset the whole product vector to zero. A similar situation occurs in the Boolean brain logic scheme with non-overlapping spike sequences in [1], where the hyperspace is additive.

This paper presents *RTW* and spike based Boolean *INBL* schemes that represent both their *H* and *L* values by non-zero noise voltages. We call them non-squeezed *INBL* schemes. They have a better potential to utilize the multidimensional logic hyperspace.

## 2. Random telegraph wave based non-squeezed logic

Let us suppose that the noise functions $H(t)$ and $L(t)$ representing the Boolean logic values are independent (orthogonal) *RTW*s with their independent random + 1 or -1 values. If we are able to realize the NOT and the AND Boolean logic functions with this system then this logic scheme is universal [1].

The NOT gate operation on an input value $X(t)$ of this type non-squeezed *INBL* logic can be realized in two different ways. The first way is additive:

$$Y = \text{NOT } X = U(t) - X(t) \qquad (1)$$

where the universe $U(t)$ is defined as

$$U(t) = H(t) + L(t) . \qquad (2)$$

The universe is not an *RTW* because its value can be -2, -1, 0, 1, and 2. It follows immediately from this definition that if $X(t) = L(t)$ then $Y(t) = H(t)$ and if $X(t) = H(t)$, then $Y(t) = L(t)$ .

The above NOT gate can be realized in hardware by a linear differential amplifier, or by a proper logic circuitry that can input multiple voltage levels (-2, -1, 0, 1, and 2) and output the proper +1/-1 levels for *RTW* generation. The last circuit would be quite involved due to the need of comparators to detect the 5 different levels of the actual universe value at the input.

The other, more advantageous way of realizing the NOT gate is via the relation:

$$Y = \text{NOT } X = X(t)H(t)L(t) \qquad (3)$$

This function is a NOT function because

$$L^2(t) = H^2(t) = 1 \qquad (4)$$

The advantage of the NOT function defined by Equation (4) is that it does not require the use of the additive universe in Equation (2), and the amplitude values remain +1 and -1 during the computation. Such a circuit can be realized by switches and binary logic operations because only binary signals exist in Equation (3).

The AND gate operation on inputs $X_1$ and $X_2$ is defined by

$$Y = X_1 \text{ AND } X_2 = \frac{1}{4}\big[H(t) - L(t)\big]\big[X_1(t) - L(t)\big]\big[X_2(t) - L(t)\big] + L(t) \qquad (5)$$

When one or more of the inputs is $L(t)$, the first term of (5) becomes 0, which makes the output $Y(t)$ equal to the remaining additive term $L(t)$. When both inputs are $H(t)$, we obtain a more complicated expression, which can be simplified by using the relation:

$$\left\{\frac{[H(t) - L(t)]}{2}\right\}^3 = \frac{H(t) - L(t)}{2} . \qquad (6)$$

This relation is easily verified by expanding the Left-Hand-Side of Equation (6) and using Equation (4) to simplify the expression.

The AND gate can be realized in hardware either by an analog circuitry (3 differential amplifiers, 2 multipliers, and one adder) or by a proper binary logic combined by analog comparator circuitry that can receive and analyze multiple discrete voltage levels and output the proper +1/-1 levels for *RTW* generation. The last circuit would be quite involved.

Since both the AND and NOT gates can be realized, we have proved that the RTW-based non-squeezed instantaneous logic scheme defined above is universal.

**3. Random spike sequence (brain) based non-squeezed logic**

In [2], a model for instantaneous, deterministic, multi-valued logic was presented, as a possible brain logic scheme. In this logic, set-theoretical operations on randomly occurring, non-overlapping (orthogonal) uniform and unipolar neural spike trains were used to construct logic superpositions and identify components in them. The key neural circuit element is the *orthon*, see Figure 1, which consists of two neurons. The idealized neurons (see Figure 1, right hand side) have two inputs, an excitatory (+) and an inhibitory (-) one. Pulses arriving at (+) will propagate to the output of the neuron, unless at the same time a pulse arrives also at the (-) input. These idealistic neurons are assumed to be free of delays [2,3]. The orthon [2] has two inputs and the $A$ and $B$ spike trains are treated as sets of spikes. The upper output provides the set-theoretical $AB$ ($A \cap B$) operation, where the overlapping spikes of $A$ and $B$ are kept and the rest are discarded. The lower output, on the other hand, provides the set theoretical $A\overline{B}$ ($A \cap \overline{B}$) operation, where the spikes of $A$ not overlapping with $B$ are kept and the rest are discarded.

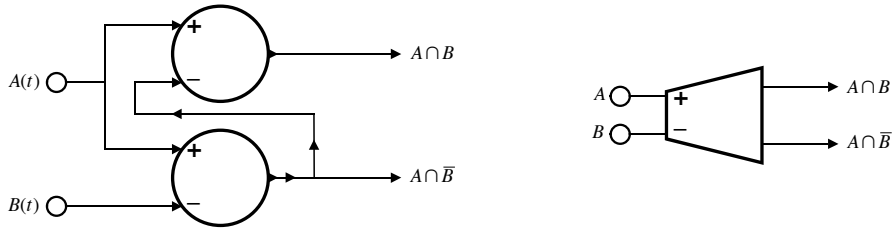

**Figure 1**. Left: the neural circuit of the *orthon* [2]. Right: its symbol.

In [1], universal Boolean logic gates for the squeezed $H(t) \neq 0$, $L(t) = 0$, version of spike-based logic were shown. Here we introduce the universal gates for the non-squeezed logic, where the logic values $H$ and $L$ are represented by the orthogonal (non-overlapping) spike trains $H(t)$ and $L(t)$, respectively, which are non-empty sequences. The orthogonality condition is

$$H(t) \cap L(t) = 0 ,  \qquad (7)$$

that is, the spikes in the two sequences do not overlap. We assume that the amplitude of a spike is 1, and that the inverse signal $\overline{X}(t)$ of $X(t)$ at time $t$ is defined by $\overline{X}(t) = 1 - X(t)$.

The NOT gate is then defined as:

$$Y(t) = \text{NOT } X = \overline{X}(t) \cap U(t) = \overline{X}(t) \cap \left[L(t) \cup H(t)\right] \qquad (8)$$

where the universe $U(t)$ is the union (sum) of $H(t)$ and $L(t)$

$$U(t) = L(t) \cup H(t) \qquad . \qquad (9)$$

This immediately implies that, for $X(t) = H(t)$, the output becomes $Y(t) = L(t)$ and, for $X(t) = L(t)$, the otput is $Y(t) = H(t)$.

The AND gate is defined as:

$$Y = X_1 \text{ AND } X_2 = [X_1(t) \cap X_2(t) \cap H(t)] \cup [X_1(t) \cap L(t)] \cup [X_2(t) \cap L(t)] \qquad (10)$$

The existence of both a NOT gate and an AND gate implies that the non-squeezed spike-based instantaneous logic scheme defined above is universal as well.

The neural hardware of the orthon-based representation of the NOT gate is shown in Figure 2.

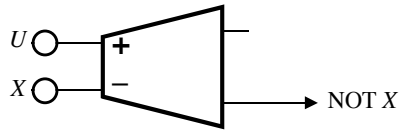

**Figure 2**. The random spike sequence based binary NOT gate utilizes the universe and the orthon element of the brain logic scheme. The orthon [2] is defined in Figure 1. The upper output is not used in the circuit.

The neural circuitry of the AND gate is shown in Figure 3. It consists of four orthons and an "adder" neuron with three excitatory inputs.

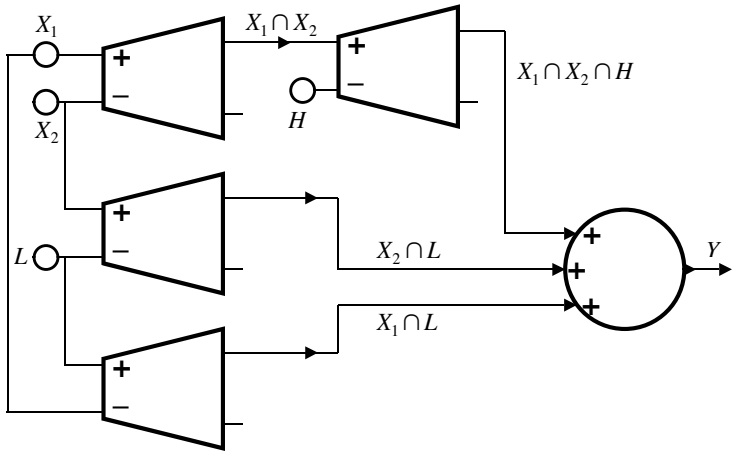

**Figure 3**. The AND gate of the brain logic. The orthon [2] is defined in Figure 1. The output neuron is summing the input pulses received at its excitatory inputs in an additive fashion (set theoretical union function).

The spike based logic scheme is clearly an instantaneous logic scheme. This is because the output is determined, without uncertainty, as soon as the first spike of the given logic value ($H$ or $L$) emerges.

## 4. Conclusions and Discussion

Two schemes of non-squeezed noise-based logic are described that are instantaneous. The lack of a need to average signals over time translates in a potentially fast on-line operation, in which inputs are operated upon as they flow in. The first scheme, which is based on Random Telegraph Waves, delivers a result of which the reliability increases as the bits arrive. The probability of ambiguity i.e., when the $H(t)$ and $L(t)$ are identical for $n$ clock steps, is $0.5^n$ [6], which translates to high reliability with relatively few bits. To achieve a reliability of $10^{-25}$ (the reliability of an idealistic CMOS gate), for example, only 83 bits are required [6].

The spike-based brain logic does even better: after only one spike, the result is established, with the remaining spikes serving as a verification of the result. This high speed of the spike-based logic takes advantage of the property that spikes in different orthogonal trains do not to overlap. Naturally, the high speed relates to the question of how long one should continue to receive spikes before cutting off a signal and switching to the next spike train. This time appears to be limited by the required reliability and, possibly, by the time period over which a certain spike train is needed to perform a particular function.

The hyperspace, in which signals can be combined before operations are conducted on them collectively, is described in more detail in [1,6]. The logic schemes presented here are more suitable for use in hyperspace, because they do not suffer from zero-multiplications in the squeezed instantaneous noise logic [1].

**Acknowledgements**

We are grateful for discussions with Sergey Bezrukov (National Institute of Health) and Sunil Khatri (Texas A&M University). LBK's visit to NICT, Japan, was supported by the NetSci subproject of NICT.